\def\pd#1#2{\frac{\partial #1}{\partial #2}}
\def\hotimes{\hat\otimes}
\def\cA{{\cal A}}
\def\cI{{\cal I}}
\def\lR{\mathbb R}

\documentclass[12pt]{revtex4}%
\usepackage{hyperref,amsfonts}
\begin{document}
\title{On linearity of separating multi-particle differential Schr\"odinger operators for identical particles}
\date{\today}
       \author{George  Svetlichny}
    \email{svetlich@mat.puc-rio.br}
    \affiliation{Departamento de Matem\'atica, \\ Pontif{\'{\i}}cia
    Universidade Cat{\'o}lica, \\ Rio de Janeiro, RJ,  Brazil}
              
   %\pacs{02.30.Jr,03.65.Ta,04.60.-m}

%\baselineskip 0pt

\begin{abstract}
\baselineskip 0pt
We show that hierarchies of differential Schr\"odinger operators for identical particles which are separating for the usual (anti-)symmetric tensor product, are necessarily linear, and offer some speculations on the source of quantum linearity.
\end{abstract}
\maketitle

\section{Introduction}

One of the properties considered in speculations about  possible fundamental non-linearities in quantum mechanics is  {\em  separation\/}, that is,  product functions evolve as product functions. Separation 
is considered  a nonlinear version of the notion of non-interaction, as then uncorrelated states remain uncorrelated under time
evolution. We show here that if separation is combined with either Fermi or Bose statistics embodied in the usual \linebreak(anti-)symmetrized  tensor product states, and if all the multi-particle Shr\"odinger operators are differential, then they are necessarily linear.

The motivation for studying  hierarchies of multi-particle non-linear Schr\"odinger equations comes from two sources: (1) intellectual speculation about possible non-linearities in quantum mechanics\cite{BBMKos}, and (2) examples   arising in 
representations of current algebras (diffeomorphism groups)\cite{HDDandGS}. We consider the second motivation compelling as current algebra representations were found to include many known linear quantum systems and to predict new ones, anyons in particular\cite{anyons}.

The non-linear theories considered still maintain that states are represented by rays in a Hilbert space, that evolution is given by a (non-linear) Schr\"odinger-type equation for the wave function, and that the modulus of the (normalized)  wave-function gives the probability density of detection. Though these assumptions can all be questioned, an important class of theories do satisfy them.

A complete analysis of separating hierarchies of  Schr\"odinger-type equations for non-identical particles was given in \cite{SVETLICHNY:GS}, however as the world is made up of bosons and fermions, the identical particle case has to be addressed.
In \cite{nlrqm} we explored the possibility of formulating a
nonlinear relativistic theory based on a nonlinear version of the
consistent histories approach to quantum mechanics. A toy model led to a set
of equation among which there were instances of a weakened form of the 
separation property for scalar bosons. This showed once more that such a 
property is fundamental for understanding any nonlinear extension of ordinary
quantum mechanics.

In \cite{SK} we showed that separating second-order differential hierarchies for identical particles are necessarily linear under various simplifying assumptions. We here prove linearity under fewer assumptions and in a more transparent fashion.  

The present result should not be  taken as an argument against non-linear quantum mechanics. As such, it would be a much weaker physical argument than the causality violation objections already raised by various authors\cite{gpl,svetBG}. Though a degree of separability is necessary to be able to isolate and observe an independent physical system, it need not be exact. Another possibility is that in non-linear theories one could conceivably form multiparticle states from states of fewer number of particles in a way other than by the usual (anti-)symmetric tensor product. In fact by using the non-linear gauge transformations of Doebner, Goldin, and Nattermann\cite{doegolnat} one can deform a linear separating hierarchy of differential Schr\"odinger operators to a non-linear hierarchy of differential Schr\"odinger separating with respect to a deformed tensor product. Whether differential hierarchies that are not equivalent to linear ones and separating with respect to deformed tensor products exist, is still to be determined.
Lastly, our results are strictly non-relativistic. 
Causal relativistic non-linear theories are seemingly hard to formulate, though they probably do exist\cite{nlrqm,bonae}. What separation implies in such a context is still to be explored. What the present result hints at is the origin of linearity about which we comment in the final section.

\section{Separation}
At time \(t\) an \(n\)-particle wave function \(\Psi\) depends on the positions \(x_1,\dots,x_n\) of each particle, where each \(x_i\in \lR^d\), \(d\) being the dimension of space, and on \(A_1,\dots,A_n\) where each \(A_i\) is an index denoting the internal degrees of freedom of each particle. Initially we assume the \(n\) particles to always belong to different species and so no permutation symmetry property is assumed of the wave-function. We use the symbol \(s=(s_1,\dots,s_n)\) as labelling the species of the particle. For initial notational ease we shall  combine the internal degrees of freedom index \(A_i\)  with the position \(x_i\) into a single symbol \(\xi_i=(x_i,A_i)\) and denote the \(n\)-tuple of such by \(\xi\).  Thus
we denote an \(n\)-particle wave function  at time \(t\) by \(\Psi(\xi,t)\).

We assume that the evolution from time \(t_1\) to time \(t_2\) of the state corresponding to  the ray with representative wave function \(\Psi(\xi,t_1)\) can be expressed by a not necessarily linear evolution operator \(E_s(t_2,t_1)\) applied to the wave-function, 
that is: 
\begin{displaymath}\Psi(\xi,t_2)=(E_s(t_2,t_1)\Psi)(\xi,t_1).
\end{displaymath}

The simple tensor product of an \(n\)- and an \(m\)-particle wave function is defined as
\begin{equation} \label{stp}
(\phi\otimes\psi)(\xi_1,\dots,\xi_n,\xi_{n+1},\dots,\xi_{n+m})=
\phi(\xi_1,\dots,\xi_n)\psi(\xi_{n+1},\dots,\xi_{n+m}).
\end{equation}

The separation property for the simple tensor product now reads: 
\begin{equation}\label{eq:stpsep}
E_s(t_2,t_1)(\Psi_1\otimes\Psi_2)= 
 E_{s_1}(t_2,t_1)(\Psi_1) \otimes E_{s_2}(t_2,t_1)(\Psi_2),
\end{equation}
where the species index \(s\) of \(\Psi\) is the concatenation of the species indices \(s_i\) of the \(\Psi_i\). Strictly speaking, since states correspond to rays and not vectors, the right-hand side should be multiplied by a complex number 
\(\gamma(t_2,t_1,s_1,s_2,\Psi_1,\Psi_2)\). To our knowledge, a full analysis of the possibility of such a factor has not been carried out. For the rest of this paper we shall assume that \(\gamma=1\), the general assumption in the literature.

Now, the world is made
of bosons and fermions and one should reconsider the separation property when one is dealing with a single species 
of identical particles. The separation property (\ref{eq:stpsep})
must then be reformulated 
with respect to the symmetric or anti-symmetric tensor product \(\phi\hat\otimes\psi\) which is the right-hand side of (\ref{stp}) symmetrized or anti-symmetrized according to either bose or fermi statistics:

\begin{equation}\label{eq:sitp} 
(\phi\hat\otimes\psi)(\xi_1,\dots,\xi_n,\xi_{n+1},\dots,\xi_{n+m})
=\frac{n!m!}{(n+m)!}\sum_{\cI}(-1)^{fp(\cI)} \phi(\xi_{i_1},\dots,\xi_{i_n})\psi(\xi_{j_1},\dots,\xi_{j_m}),
\end{equation}

where  \(\cI=(i_1,\dots,i_n)\) are \(n\)
numbers from \(\{1,\dots,n+m\}\), in ascending order, \((j_1,\dots,j_m)\) the
complementary numbers, also in ascending order, \(f\) is the {\em  Fermi number\/} \(0\) for bosons and \(1\) for fermions, and \(p(\cI)\) is the parity
(\(0\) for even and \(1\) for odd) of the permutation \((1,\dots,n+m)\mapsto (i_1,\dots,i_n,j_1,\dots,j_m)\). We have taken into account that both \(\phi\) and \(\psi\) are either symmetric or antisymmetric with respect to permutations of their arguments. The normalizing factor makes the product associative and the map \(\phi\otimes\psi\mapsto\phi\hotimes\psi\) into a projection. For the identical particle case, the species symbol \(s\) reduces just to the particle number \(n\).

If we pass to the generators of the evolution operators
\begin{displaymath}  H_s(t)=\left.\frac1i\frac{\partial}{\partial t_2}E_s(t_2,t_1)\right|_{t_2=t_1=t}
\end{displaymath}
then the separation property (\ref{eq:stpsep}) (under the assumption that \(\gamma=1\)) becomes:
\begin{equation}\label{eq:stender}
H_s(\Psi_1\otimes\Psi_2)
= H_{s_1}(\Psi_1)\otimes\Psi_2+\Psi_1\otimes
H_{s_2}(\Psi_2),
\end{equation}
where for notational simplicity we have suppressed  indicating the \(t\) dependence of the \(H\)'s. This relation (which we called {\em  tensor derivation\/}) was fully analyzed in \cite{SVETLICHNY:GS}. Canonical
decompositions and constructions were also presented.

An (anti-)symmetric tensor derivation would be a hierarchy of operators \(H_n\) that
satisfies  (\ref{eq:stender}) with \(\hat\otimes\) instead of
\(\otimes\). 
One does not have a classification of these as one has for ordinary tensor
derivations as given in \cite{SVETLICHNY:GS}. It seems that the conditions to
be a tensor derivation in the \linebreak(anti-)symmetric case is rather stringent, and as we
shall now see, in the case of differential operators, implies linearity. 

It now becomes convenient to disentangle the space-coordinate \(x\) and the internal degree of freedom index \(A\).
Our one-particle wave function will thus be denoted by \(\psi^A(x)\) with the index as a superscript for convenience. Multi-particle wave function will carry multiple indices in the usual way. The possibly non-linear operators of the tensor derivation will be assumed to depend on the real and imaginary parts of the wave function in an independent fashion, though, to simplify notation, this is not denoted explicitly. Likewise, for notational ease, internal degree of freedom indices will be suppressed when no confusion can arise.

We shall use a multi-index notation for partial derivatives. Given a function \(u(x_1,\dots,x_n)\) and \(I=(i_1,\dots,i_n)\) an \(n\)-tuple of non-negative integers, we denote by \(|I|\) the sum \(i_1+\cdots+i_n\) and by \(u_I\) the partial derivative
\begin{displaymath}
\partial_I u=\frac{\partial^{|I|}u}{\partial x_1^{i_1}\cdots\partial x_n^{i_n}}.
\end{displaymath}
For the case of a function \(u(x,y)\) of two variables we write \(u_{I,J}\) for \(I\) differentiations with respect to \(x\), and \(J\) with respect to \(y\).

 Let us consider  possibly nonlinear differential
operators of any order (dependence on time can be construed as simply dependence on a
parameter). Such a two-particle operator has the form \(H(x,y,\phi^{AB}_{I,J}(x,y))\). Introducing variable names for
the arguments of \(H\), we write \(H(x,y,a^{AB}_{I,J})\).
When \(\phi\) is constrained to be an (anti-)symmetrized product 
\begin{displaymath}\phi^{AB}(x,y) =
\frac12(\alpha^A(x)\beta^B(y)+(-1)^f\beta^A(x)\alpha^B(y)),
\end{displaymath} 
then the arguments of $H$ are
constrained to take on values of the form.
\begin{equation}\label{svetlichny:ra}
a^{AB}_{I,J}=\frac12(\alpha^A_I\tilde\beta^B_J+(-1)^f\beta^A_I\tilde\alpha^B_J).
\end{equation}
Here quantities without the tilde are derivatives evaluated at \(x\) and those with, at \(y\).
The quantities on the right-hand sides:  $\alpha^A_I, \beta^A_I,
\tilde\alpha^B_J, \tilde\beta^B_J$, which we
shall call the \(\alpha\beta\)-quantities, can be given, by Borel's lemma,
arbitrary complex values by an appropriate choice of the points \(x\) and
\(y\) and functions \(\alpha\) and \(\beta\). Denote the right-hand sides of
the above equations by  \(\hat a^{AB}_{I,J}\). 

 The separability condition for the symmetrized
tensor product now reads:
\begin{equation}\label{svetlichny:2sep}
2H^{AB}_2(x,y, \hat a_{I,J})  =
H^A_1(x,\alpha_I)\tilde\beta^B_{0} + 
\alpha^A_{0}H^B_1(y ,\tilde\beta_J)+
(-1)^fH^A_1(x, \beta_I)\tilde\alpha^B_{0}+
(-1)^f\beta^A_{0}H^B_1(y, \tilde\alpha_J). 
\end{equation}

Now we come to the main point: in the space of the \(\alpha\beta\)-quantities there are flows that leave  \(\hat a_{I,J}\) invariant, and so must leave the right-hand side of (\ref{svetlichny:2sep}) invariant. This leads to linearity.

\section{Proof of linearity}

One easily sees that the following transformations leave the \(\alpha\beta\)-quantities invariant:
\begin{eqnarray}\label{scale}\nonumber
\alpha_I^A\mapsto s\alpha_I^A,&& \tilde\beta_J^B\mapsto s^{-1} \tilde\beta_J^B;\\ \label{shift}
\alpha_I^A\mapsto \alpha_I^A+s\beta_I^A, &&\tilde\alpha_J^B\mapsto \tilde\alpha_J^B-s(-1)^f\tilde\beta_J^B; 
\end{eqnarray}
and the same with \(\alpha\) and \(\beta\) interchanged. Symmetry (\ref{shift}) is enough to force linearity. 

Note that \(s\) is a {\em  complex\/} parameter, which means that the real and imaginary parts of the quantities undergo separate transformations. As a result, the right-hand side of (\ref{svetlichny:2sep}) has to be annihilated by
the vector field corresponding to (\ref{shift}): 
\begin{equation}\label{abiflow}
\sum_{C,I}\left(\beta^C_I\pd{}{\alpha^C_I}-(-1)^f\tilde\beta^C_I\pd{}{\tilde\alpha^C_I}\right),
\end{equation}
where by \( \partial/\partial \alpha^C_I\) we mean the usual convention \( (1/2)\left(\partial/\partial \hbox{Re}\,\alpha^C_I-i\partial/\partial\hbox{Im}\alpha^C_I\right)\) and similarly for the other partial derivative.

Applying now (\ref{abiflow}) to the right-hand side of (\ref{svetlichny:2sep}), we get:
\begin{displaymath}
\left[\sum_{C,I}\beta_I^C\pd{H_1^A}{{\alpha^C_I}}(x,\alpha)-H_1^A(x,\beta)\right]\tilde \beta^B_0-\beta^A_0\left[\sum_{C,I}{{\tilde\beta}_I^C} \pd{H_1^B}{{{\tilde\alpha}^C_I}}(y,\tilde\alpha)-H_1(y,\tilde \beta)\right]=0.
\end{displaymath}
Now the \(\alpha\beta\)-quantities can be chosen arbitrarily and generically we have \(\beta_0^A\neq 0\) and \(\tilde\beta_0^B\neq 0\) for all \(A\) and \(B\) and so generically 
\begin{displaymath}
\frac1{\beta^A_0}\left[\sum_{C,I}\beta_I^C
\pd{H_1^A}{{\alpha^C_I}}(x,\alpha)-H_1^A(x,\beta)\right]=
 \frac1{\tilde \beta^B_0}\left[\sum_{C,I}{{\tilde\beta}_I^C} \pd{H_1^B}{{{\tilde\alpha}^C_I}}(y,\tilde\alpha)-H_1(y,\tilde \beta)\right].
\end{displaymath}
Since both sides depend on different sets of variables, each side is a constant \(k\) and we now have: 
\begin{displaymath}
\sum_{C,I}\beta_I^C\pd{H_1^A}{{\alpha^C_I}}(x,\alpha)-H_1^A(x,\beta)=k\beta^A_0.
\end{displaymath}
Fixing \(\alpha\) this equation states that \(H_1(x,\beta)\) is a linear function of \(\beta\) with coefficients depending on \(x\). We have thus shown:

{\bf Lemma:} {\sl
In an  (anti-)symmetric tensor derivation in which the one-particle and two-particle operators are differential, the one-particle operator is necessarily linear. }

To show the whole hierarchy is linear we procede as in 
 \cite{SK}. An \(N\)-particle wave-function for particles in \(\lR^d\) can be viewed as a one-particle wave-function for particles (call them {\em
conglomerate\/} particles) in \(\lR^{Nd}\). Consider
the separating property for a
\(2N\)-particle operator acting on an (anti-)symmetrized tensor product of two
\(N\)-particle wave-functions, reinterpreted now as a separating property for operators acting on the wave-functions of  two and one conglomerate particles. 
The only difference in relation to what we have already done, is the permutation symmetry of conglomerate particles.
Let \(\phi(x_1,\dots,x_N)\) and \(\psi(y_1,\dots,y_N)\)
be two properly (anti-)symmetric \(N\)-particle wave-functions. One has using the conventions of (\ref{eq:sitp}):
%\begin{widetext}
\begin{equation} \label{svetlichny:2congl}(\phi\hat\otimes\psi)^{\cA}(x_1,\dots,x_{2N})
=\frac{N!^2}{(2N)!}\sum_{\cI}(-1)^{fp(I)} \phi^{\cA_I}(x_{i_1},\dots,x_{i_N})\psi^{\cA_J}(x_{j_1},\dots,x_{j_N}),
\end{equation}
%\end{widetext}
where \(\cA=(A_1,\dots,A_{2N})\), \(\cA_I=(A_{i_1},\dots,A_{i_N})\), and \(\cA_J=(A_{j_1},\dots,A_{j_N})\) are internal degree of freedom indices. For (\ref{svetlichny:2congl}) the possible values that one
can attribute to the wave-function and its derivatives at a point is now more
complicated than that given by (\ref{svetlichny:ra}), but since by an appropriate choice of
coordinates and an appeal to Borel's lemma we can again use   (\ref{svetlichny:ra}) as a particular case for two 
conglomerate particles, the only differences being the change of the combinatorial factor \(1/2\) to \(N!^2/(2N)!\)  and the possibility that the  factor \((-1)^f\) may be absent even in the Fermi case. These differences are non-essential to the derivation, and repeating the argument presented above for the
two-particle case we  see that the operator for  one conglomerate particle
must be linear and so the \(N\)-particle operator must be linear. With this
the whole hierarchy must be linear. We thus have:

{\bf Theorem:} {\sl An (anti-)symmetric tensor derivation in which  all multiparticle operators are  differential, is  necessarily linear.} 
\section{Comments on the origin of quantum linearity}

Our view on quantum-mechanical linearity is that it is an emergent feature of the world that arises along with the manifold structure of space-time from some more fundamental pre-geometric reality. Thus questions of (non)linearity should be joined with the general quantum gravity program. Previous clues in this direction are provided by (1) the apparent connections between linearity and the causal structure of space-time\cite{svetBG,cover} and by (2) the difficulty of incorporating internal degrees of freedom, such as spin, in separating non-linear theories, requiring new multi-particle effects at every particle number\cite{tandor}. We consider the present result as another such clue, linking linearity to the statistics of identical particles and the possibility of independently evolving systems. 

The emergent view of linearity is also supported by the present extremely small experimental bounds on possible non-linear effects, the suppression factor being about \(10^{-20}\)\cite{nlexp1}. If linearity is emergent, experimental evidence would be hard to come by. There is however the possibility  that ultra-high-energy cosmic rays actually do probe the hypothetically non-linear pre-geometric regime\cite{cosmicrays}.  The possible role of non-linearities on the Planck scale has also been considered by T.~P.~Singh\cite{singh}, and by N.~E.~Mavromatos and R.~J.~Szabo\cite{mavsz}.

\subsection*{Acknowledgements}
This research was partially supported by the Conselho Nacional de Desenvolvimento Scient\'{\i}fico e Tecnol\'ogico (CNPq).


\begin{thebibliography}{xx}
\bibitem{BBMKos}M.~D.~Kostin,   J.~Chem.~Phys.  {\bf 57}, 3589 (1972); I.~Bialynicki-Birula and J.~Mycielski,
  Ann.~Phys.~(N.Y.) {\bf 100}, 62 (1976); S.~Weinberg, Phys.~Rev.~Lett. {\bf 62}, 485 (1989); Ann.~Phys.~(N.Y.) {\bf 194}, 336 (1989).

\bibitem{HDDandGS} H.-D.~Doebner and G.~A.~Goldin,
  Phys.~Lett.~A  {\bf 162}, 397 (1992);
G.~A.~Goldin,
  Int.~J.~Mod.~Phys.  {\bf B6}, 1905 (1992);
H.-D.~Doebner and G.~A.~Goldin,
in{\em  Proceedings of the XIX\(^{\rm th}\) International Conference
on Group Theoretical
Methods in Physics, Salamanca, June 29 - July 4, 1992\/},
edited by J.~M.~Guilarte, M.~A.~del~Olmo, and
M.~Santander, {\em Annales de Fisica, Monografias,
Vol. II\/} (CIEMAT/RSEF, Madrid, 1993), 442.


\bibitem{anyons}G.~A.~Goldin and  D.~H.~Sharp    Phys.~ Rev.~D  {\bf 28}, 830 (1983).

\bibitem{SVETLICHNY:GS}G.~A.~Goldin and G.~Svetlichny,  J.~Math.~Phys.
{\bf 35}, 3322 (1994).

\bibitem{nlrqm}G.~Svetlichny, in
{\it Proceedings of the Second International Conference
``Symmetry in Nonlinear Mathematical Physics.
Memorial Prof. W. Fushchych Conference"\/} edited by M.~Shkil, A.~Nikitin, and V.~Boyko  (Mathematics Institute of the National Academy of Sciences of Ukraine, Kyiv, 1997), Vol.\ 2, p.\ 262. 

\bibitem{SK} G.~Svetlichny,   in 
{\em  Proceedings of the Fourth International Conference
``Symmetry in Nonlinear Mathematical Physics"\/}, edited by A.~G.~Nikitin, V.~M.~Boyko and R.~O.~Popovich  (Mathematics Institute of the National Academy of Sciences of Ukraine, Kyiv,  2002) Vol.\ 2, p.\ 691.

\bibitem{gpl} N.~Gisin,   Helv.~Phys.~Acta  {\bf 62}, 363 (1989);
J.~Polchinski,   Phys.~Rev.~Lett.  {\bf 66}, 397 (1992);
W.~Luecke, ``Nonlocality in Nonlinear Quantum Mechanics" [quant-ph/9904016].

\bibitem{svetBG} G.~Svetlichny,    Found.~Phys. {\bf 28}, 131 (1998) [quant-ph/9511002];
C.~Simon,  V.~Bu\v{z}ek and N.~Gisin,   Phys.~Rev.~Lett.  {\bf 87}, 170405 (2001) [quant-ph/0102125].

\bibitem{doegolnat} H.-D.~Doebner, G.~A.~Goldin, Phys.~Rev.~A {\bf 54}, 3764 (1996); H.-D.~Doebner, G.~A.~Goldin, and P.~Nattermann, J.~Math.~Phys. {\bf 40}, 49 (1999) 

\bibitem{bonae} P.~Bona in {\em Proceedings of Conference: New Insights in Quantum Mechanics, Goslar, Aug. 31\({}^{\rm st}\)-Sept. 4\({}^{\rm th}\) 1998\/} (World Scientific 1999) [quant-ph/9910011];
S.~Gheorghiu-Svirschevski, ``A General Framework for Nonlinear Quantum Dynamics" [quant-ph/0207042];
A.~Kent, ``Nonlinearity without Superluminality" [quant-ph/0204106].

\bibitem{cover}G.~Svetlichny, Found.~Phys. {\bf 30}, 1819 (2000)
[quant-ph/9912099].
\bibitem{tandor}G.~Svetlichny,    Journal of Nonlinear Mathematical
Physics {\bf 2}, 2 (1995) [quant-ph/9906117].

\bibitem{nlexp1} J.~J.~Bollinger, D.~J.~Heinzen, W.~M.~Itano, S.~L. Gilbert, and 
D.~J.~Wineland,   Phys.~Rev.~Lett.  \textbf{63}, 1031 (1989);
R.~L.~Walsworth, I.~F.~Silvera, E.~M.~Mattison and R.~F.~C.~Vessot,   {\sl ibid.\/}  \textbf{64}, 2599 (1990);
T.~E.~Chupp and R.~J.~Hoare,   {\sl ibid.\/}   \textbf{64}, 2261 (1990);
 P.~K.~Majumder, B.~J.~Venema, S.~K.~Lamoreaux, B.~R.~Heckel, and 
E.~N.~Fortson,  {\sl ibid.\/}  \textbf{65}, 2931 (1990);
 F.~Benatti and R.~Floreanini,   Phys.~Lett.  \textbf{B389}, 100 (1996); \textbf{B451}, 422 (1999).

\bibitem{cosmicrays}G.~Svetlichny, ``Non-linear quantum mechanics and  high energy cosmic rays" [hep-th/0305100].

\bibitem{singh}T.~P.~Singh, ``Quantum mechanics without spacetime III - A proposal for a non-linear Schrodinger equation -" [gr-qc/0306110].

\bibitem{mavsz} N.~E.~Mavromatos and R.~J.~Szabo,  Int.\ J.\ Mod.\ Phys.\ \textbf{A16}, 209 (2001)  [hep-th/9909129].

%
\end{thebibliography}
\end{document}